\documentclass[12pt,reqno]{amsart}
\RequirePackage{multirow}
\usepackage{xifthen,setspace,multirow}
\usepackage{diagbox,graphicx}
\usepackage{caption,subcaption}
\usepackage{float}
\usepackage[margin=3cm]{geometry}
\usepackage{bbm}
\usepackage{xcolor}
\usepackage{enumitem}
\usepackage{bm}
\usepackage[linesnumbered,ruled,vlined]{algorithm2e}
\usepackage{hyperref}
\usepackage{esvect}
\usepackage{amsmath,amssymb,mathrsfs} 
\usepackage{lipsum}  

\newtheorem{theorem}{Theorem}

\newtheorem{proposition}{Proposition}
\newtheorem{lem}{Lemma}
\newtheorem{assumption}{Assumption}
\newtheorem{condition}{Condition}

\newcommand{\vertiii}[1]{{\left\vert\kern-0.25ex\left\vert\kern-0.25ex\left\vert #1 
    \right\vert\kern-0.25ex\right\vert\kern-0.25ex\right\vert}}

\renewcommand{\leq}{\leqslant}
\renewcommand{\geq}{\geqslant}

\newcommand{\ris}{\hat{\bm J}_{r,I}}
\newcommand{\sech}{\mathrm{sech}}
\newcommand{\rpl}{\hat{\bm J}_{r,P}}
\newcommand{\bpk}{{\binom{p-1}{k-1}}}

\newcommand{\p}{{\mathbb{P}}}

\newcommand{\eS}{\mathcal{S}}

\allowdisplaybreaks

\title[Tensor Learning in Ising Models]{Interaction Screening and Pseudolikelihood Approaches for Tensor Learning in Ising Models}

\author[Liu]{Tianyu Liu}
\thanks{The first two authors contributed equally to the paper.}
\address{email {\tt tianyu.liu@u.nus.edu}}

\author[Mukherjee]{Somabha Mukherjee}
\address{email {\tt somabha@nus.edu.sg}}

\begin{document}

\begin{abstract} 
In this paper, we study two well known methods of Ising structure learning, namely the pseudolikelihood approach and the interaction screening approach, in the context of tensor recovery in $k$-spin Ising models. We show that both these approaches, with proper regularization, retrieve the underlying hypernetwork structure using a sample size logarithmic in the number of network nodes, and exponential in the maximum interaction strength and maximum node-degree. We also track down the exact dependence of the rate of tensor recovery on the interaction order $k$, that is allowed to grow with the number of samples and nodes, for both the approaches. We then provide a comparative discussion of the performance of the two approaches based on simulation studies, which also demonstrates the exponential dependence of
the tensor recovery rate on the maximum coupling strength. Our tensor recovery methods are then applied on gene data taken from the Curated Microarray Database (CuMiDa), where we focussed on understanding the important genes related to hepatocellular carcinoma.
\end{abstract}



	\maketitle

\section{Introduction}\label{int}
The $k$-spin Ising model \cite{barra, oliveira, gardner, bovier, mukherjeeestimation_b} is a discrete exponential family for modeling dependent binary ($\pm 1$-valued) data exhibiting multi-body interactions, which can be thought of as taking place along the hyperedges of a $k$-uniform hypergraph. It is a generalization of the classical 2-spin Ising model, that was originally introduced by physicists for modelling ferromagnetism \cite{ising}, and has since then found extensive applications in diverse areas such as social sciences, image processing, computational biology, neural networks,   spatial statistics and election forecasting \cite{spatial,cd_trees,geman_graffinge,disease,neural,innovations,elhj1,isinggenapl5}. However, pairwise interactions are often not adequate to capture all the complex dependencies that arise in real-world network structures, such as peer-group effects and multi-atomic interactions on crystal surfaces, which necessitates the use of higher order tensor Ising models for modeling such complex relational frameworks. The multi-body
interactions in tensor Ising models are captured by an interaction tensor, which in many practical cases, forms the adjacency tensor of an underlying hypergraph, whose nodes
are thought of as being sites for binary-valued variables or spins, that interact with each other along the hyperedges of this hypergraph. A fundamental problem in the area of Ising model learning is to recover this underlying interaction structure given access to multiple samples from this model, on which a significant amount of literature has developed for the classical 2-spin Ising models, during the past two decades (see,
for example, \cite{structure_learning,bresler,discrete_tree,graphical_models_algorithmic,wainwright,graphical_models_binary,lokhov}). Variants of the complete structure learning problem have also experienced growing interest over the past few years, some related notable works being graph property testing by Neykov et al.  \cite{high_tempferro,neykovliu_property}, identity and independence testing by Daskalakis et al. \cite{cd_testing}, and Ising model learning by Devroye et al. \cite{luc}.

Ravikumar et al. \cite{wainwright} implemented an $\ell_1$-penalized nodewise logistic regression approach to recover the signed neighborhood of each node of the underlying Ising graph. However, as pointed out in \cite{montanarig}, the \textit{incoherence condition} assumed in \cite{wainwright} is $\mathrm{NP}$-hard to check for general non-ferromagnetic Ising models, and is computationally expensive in the ferromagnetic case, too. Bresler \cite{bresler} proposed a greedy algorithm for exact recovery of an Ising interaction matrix using a sample of size logarithmic in the network size, that has runtime complexity much lower than the $\ell_1$-regularized nodewise logistic regression approach in \cite{wainwright}. However, practically speaking, the applicability of this method does not extend much beyond bounded-degree networks, since the dependence of the sample size on the maximum node degree is double exponential in nature. For related works on structure learning in antiferromagnetic and tree-structured Ising models, see \cite{bresler1, bresler2}. 
Vuffray et al. \cite{vuffrayint} proposed a quasi-quartic time algorithm based on \textit{interaction screeing}, that also works on sample sizes logarithmic in the network size, but exponential in the order of the maximum node degree and the maximum interaction strength. This scaling, unlike the algorithm in \cite{bresler}, matches the information theoretic lower bound of \cite{graphical_models_binary}. As an added advantage, unlike the approach in \cite{bresler}, the interaction screening approach in \cite{vuffrayint} does not require prior knowledge of model parameters, such as the maximum interaction strength and largest node degree, which may not always be obtainable.

In spite of the presence of a significant amount of literature on matrix structure learning in classical $2$-spin Ising models, to the best of our knowledge, analogous literature for tensor recovery in $k$-spin Ising models is extremely scarce, if at all existent. Only recently, Liu at al. \cite{liu} extended the technique and results in \cite{wainwright} to tensor recovery in $k$-spin Ising models. In this paper, we study the interaction screening \cite{vuffrayint} and the regularized pseudolikelihood methods for tensor recovery in $k$-spin Ising models. As was observed in \cite{montanarig}, the regularized pseudolikelihood algorithm performs very poorly for high values of the maximum coupling strength, despite working with a sample size much larger than the logarithm of the number of nodes, and irrespective of the choice of the LASSO tuning parameter. Further, the threshold for the maximum coupling strength, above which this poor performance emerges, is also very near to the critical point for the Ising model on a randomly diluted grid (see \cite{montanarig}). The dependence of the minimum sample size on the maximum coupling strength was not made explicit in \cite{wainwright} or \cite{liu}, and was incorporated within the proportionality constant. Inspired by \cite{vuffrayint}, in this paper, we work out this dependency, and show that a sample size exponential in the maximum coupling strength guarantees consistent tensor recovery. This is a plausible explanation behind the worse behavior of the regularized estimators at low temperatures (high coupling strengths). We also track down the dependence of the minimum sample size and rate of convergence on the order $k$ of interaction. In the following subsection, we describe the $k$-spin Ising model and the tensor recovery problem in more details.

\subsection{The Tensor Recovery Problem}
The $k$-spin or $k$-tensor Ising model is a probability distribution on the set $\{-1,1\}^p$, defined as:
\begin{equation}\label{modeldef}
    \p_{\bm J}(\bm x) := \frac{1}{Z(\bm J)}e^{H(\bm x)}\quad(\bm x \in \{-1,1\}^p)
\end{equation}
where $\bm J := ((J_{r_1,\ldots,r_k}))_{(r_1,\ldots,r_k)\in [p]^k}$ denotes a $k$-fold tensor with $[p] := \{1,\ldots,p\}$, $$H(\bm x) := \sum_{(r_1,\ldots,r_k)\in [p]^k} J_{r_1,\ldots,r_k} x_{r_1}\ldots x_{r_k},$$ and $Z(\bm J)$ is a normalizing constant required to ensure that $\p_{\bm J}$ is a valid probability distribution.
Hereafter, we will assume that the tensor $\bm J$ satisfies the following properties:
\begin{enumerate}
    \item $\bm J$ is symmetric, i.e., $J_{r_1,...,r_k}= J_{r_{\sigma (1)},...,r_{\sigma(k)}}$ for every $(r_1,\ldots,r_k)\in [p]^k$ and every permutation $\sigma$ of $\{1,...,k\}$,
    \item $\bm J$ has zeros on the \textit{diagonals}, i.e., $\bm J_{r_1...r_k}=0$, if $r_s=r_t$ for some $1\leqslant s < t \leqslant k$.
\end{enumerate}
Also, throughout the paper, we will denote the maximum entry of $\bm J$ by $\beta$, i.e. 
$$\beta := \max_{(r_1,\ldots,r_k)\in [p]^k} J_{r_1,\ldots,r_k}~.$$
The aim of this paper is to recover the unknown tensor $\bm J$ given access to samples $\bm x^{(1)},\bm x^{(2)},\ldots, \bm x^{(n)}$ from this model. Based on the works \cite{lokhov,vuffrayint} in the context of Ising matrix recovery, in this paper, we analyze the performance of two approaches, namely the  \textit{regularized interaction screening estimator} and the \textit{regularized pseudolikelihood estimator} on tensor recovery in higher-order Ising models. 

The regularized interaction screening estimator (RISE) of the neighborhood of a particular node $r\in [p]$ is defined as:
\begin{equation}\label{lasso_rise}
   \hat{\bm J}_{r,I} := \arg\min_{\bm J_r \in \mathbb{R}^{\bpk}} \eS(\bm J_r;\mathfrak{X}^n) + \lambda \|\bm J_r\|_1
\end{equation}
where $\bm J_r := ((J_{r,r_1,\ldots,r_{k-1}}))_{ (r_1,\ldots,r_{k-1})\in T_r}$ and
$$\eS(\bm J_r;\mathfrak{X}^n) := \frac{1}{n} \sum_{i=1}^n \exp \left(-k x^{(i)}_r m_r(\bm x^{(i)})\right),$$
    with $m_r(\bm x) := \sum_{(r_1,\ldots,r_{k-1}) \in [p]^{k-1}} J_{r,r_1,\ldots,r_{k-1}} x_{r_1}\ldots x_{r_{k-1}}$ and $$T_r := \{(r_1,\ldots,r_{k-1}) \in ([p]\setminus \{r\})^{k-1}: 1\le r_1<\ldots <r_{k-1} \le p\}.$$
Note that the objective function in \eqref{lasso_rise} is convex, and hence, \eqref{lasso_rise} is a convex optimization problem. 

On the other hand, the regularized pseudolikelihood estimator (RPLE) of the neighborhood of the node $r\in [p]$ is defined as:

\begin{equation}\label{lasso9}
    \hat{\bm J}_{r,P} := \arg \min_{\bm J_r \in \mathbb{R}^{\bpk}} \ell(\bm J_r;\mathfrak{X}^n) + \lambda \|\bm J_r\|_1
\end{equation}
where 
$$\ell(\bm J_r;\mathfrak{X}^n) := -\frac{1}{n} \sum_{i=1}^n \log \p_{\bm J}(x_r^{(i)}| \bm x_{\setminus r}^{(i)})$$
and $\bm x_{\setminus r}^{(i)} := (x_t^{(i)})_{t\ne r}$. A straightforward computation shows that:
$$\p_{\bm J}(x_r| \bm x_{\setminus r}) = \frac{\exp(k x_r m_r(\bm x))}{2\cosh(k x_r m_r(\bm x))}$$
and hence, one has:
$$\ell(\bm J_r;\mathfrak{X}^n) = -\frac{1}{n}\sum_{i=1}^n \left\{kx_r^{(i)} m_r(\bm x^{(i)}) - \log \cosh(k x_r^{(i)} m_r(\bm x^{(i)})) - \log 2\right\}$$

In this paper, we study the performances of both the tensor-structure learners RISE $\ris$ and RPLE $\rpl$. In Section \ref{sec:rpl}, we prove rates of consistency of the RISE and the RPLE, and show explicit dependence of these rates and the minimum sample size requirement on the maximum coupling strength $\beta$ and the tensor interaction factor $k$.

\subsection{Related Work on Parametric Inference in Ising Models}
A closely related problem is the task of infering the \textit{inverse temperature} and the \textit{external magnetic field} parameters from the Ising model:
$$\p(\bm x) \propto e^{\beta \bm x^\top \bm J \bm x + h\sum_{i=1}^n x_i}\quad(x\in \{-1,1\}^p)$$
given access to a single sample from such a model. There has been a flurry of works in the past two decades in the area of parametric inference in Ising models, starting with the seminal paper due to \cite{chatterjee}, who, inspired by \cite{besag_nl, besag_lattice}, first applied the pseudolikelihood approach for parameter estimation in general spin-glass models, and showed $\sqrt{p}$-consistency of the maximum pseudolikelihood estimator in this model with $h=0$, at low temperatures. This was followed by improved results on the rate of consistency and joint estimation of $(\beta,h)$ in \cite{BM16} and \cite{pg_sm}. Recently, some of these results were also extended to tensor Ising models in \cite{mukherjeeestimation_b} and \cite{fluctmj}

\subsection{Structure of the Paper}
The rest of the paper is organized as follows. In Section \ref{sec:rpl}, we state the main results of this paper. We begin by discussing about a general theory of $\ell^1$-penalized $M$-estimators from \cite{nega}, that will help us prove the main results in this paper. In the same section, we use this general theory to prove theoretical guarantees for the RPLE and the RISE. In Section \ref{sec:simul}, we provide some simulation studies to demonstrate the comparative performance of these two approaches. In Section \ref{sec:realdata}, we implement these methods on a real-life gene data taken from the Curated Microarray Database (CuMiDa), with the aim of discovering the important genes related to hepatocellular carcinoma. Proofs of some technical results that are not included in the main paper are given in the appendix.

\section{Main Results}\label{sec:rpl}
In this section, we state and prove the main results of this paper. To begin with, we sketch a general theory about $\ell_1$-penalized $M$-estimators from \cite{nega}, that is crucial in proving our main results,  
\subsection{A general theory of $\ell_1$-penalized $M$-estimators}
We use a framework from \cite{nega} for general $\ell_1$-regularized $M$-estimators to establish that our method works consistently. It turns out that imposing just the following two conditions on the loss function is enough to control the error of the $\ell_1$-regularized $M$-estimator:

$$\hat{\bm J}_r := \arg \min_{\bm J_r \in \mathbb{R}^{\bpk}} \mathcal{L}(\bm J_r;\mathfrak{X}^n) + \lambda \|\bm J_r\|_1$$
for a general convex and differentiable loss function $\mathcal{L}$.

\begin{condition}\label{cond1}
    Let $\lambda$ be the $\ell_1$-penalty parameter. Then, the gradient of the objective function at the true neighborhood $\bm J_r$ satisfies:
    \begin{equation*}
         2\|\nabla \mathcal{L}(\bm J_r;\mathfrak{X}^n)\|_{\infty} \leq \lambda
    \end{equation*}
\end{condition}
\textcolor{black}{Condition 1 ensures that if the maximum degree of the hypergraph with (weighted) adjacency $\bm J$ is $d$, then the difference $\eta_r:= \hat{\bm J_r}-\bm J_r$ lies within the two-sided cone:
   \begin{equation}\label{condecnd2}
       K:=\left\{\eta\in \mathbb{R}^{p-1 \choose k-1} ~\Big |~\|\eta\|_1 \leq 4\|\eta_S\|_1\right\},
    \end{equation}
    where $S$ denotes the indices of all non-zero entries of $\bm J_r$. Note that \eqref{condecnd2} follows from Lemma 1 and Example 1 in \cite{nega}.}

\begin{condition}[Restricted Strong Convexity]\label{cond2}
    There exists $R>0$ such that for all $\eta_r\in K$ with $\|\eta_r\|_2 \leq R$, there exists a constant $\kappa >0$, such that
    \begin{equation*}
        \mathcal{L}(\bm J_r + \eta_r;\mathfrak{X}^n) - \mathcal{L}(\bm J_r;\mathfrak{X}^n) - \langle \nabla \mathcal{L}(\bm J_r;\mathfrak{X}^n), \eta_r\rangle \geq \kappa \|\eta_r\|_2^2.
    \end{equation*}
\end{condition}
The second condition guarantees that the loss function is strongly convex in a conically restricted neighborhood of $\mathbb R^{p-1 \choose k-1}$. We will verify Conditions \ref{cond1} and \ref{cond2} for the RISE and the RPLE later. With these two conditions and the above assumption in hand, we can finally proceed to bound the error of estimation of the neighborhood parameter $\bm J_r$. The following proposition serves as the fundamental result begind our main theorem about the theoretical guarantee of the RPLE:
 
\begin{proposition}\label{prop11}
      Under Conditions \ref{cond1} and \ref{cond2} with $R > 3\sqrt{d}\lambda/\kappa$, we have:
    \begin{equation*}
        \|\hat{\bm J}_{r} - \bm J_r\|_2 \leq \frac{3\lambda\sqrt{d}}{\kappa}.
    \end{equation*}
    \end{proposition}

The following restricted eigenvalue condition on the tensor covariance structure is also necessary for our analysis: 

\begin{assumption}\label{asm1}
    Let $Q := \mathbb{E}[\bm X_{\cdot r} \bm X_{\cdot r}^\top]$ where $\bm X_{\cdot r} := (X_{r_1}\ldots X_{r_{k-1}})_{(r_1,\ldots,r_{k-1}) \in T_r}$. Assume that there exists a constant $\alpha > 0$, such that:
    $$\inf_{\bm v\in K\setminus \{\boldsymbol{0}\}}\frac{\bm v^\top Q\bm v}{\|\bm v\|_2^2} \ge \alpha~.$$    
\end{assumption}
Note that, a sufficient condition for Assumption \ref{asm1} to hold, is that the minimum eigenvalue of $Q$ is bounded below by $\alpha$. With this structure in hand, we are finally ready to prove the theoretical guarantees of the RISE and RPLE.

\subsection{The Regularized Pseudo-Likelihood Estimator}
In this section, we analyze the performance of the RPLE. The main result about the rate of convergence of the RPLE is stated below: 

\begin{theorem}\label{thm1}
 Suppose that $d$ denotes the maximum degree of the hypernetwork with adjacency $\bm J$, and the regularization parameter $\lambda$ is chosen as:
\begin{equation*}\label{reg}
    \lambda :=  4\sqrt{2} k! \sqrt{\frac{\log\left(4\bpk/\varepsilon\right)}{n}}.
\end{equation*}
Then, there exist constants $M_1,M_2>0$, such that for every node $r\in V$ and any $\varepsilon \in (0,1)$, if
\begin{equation*}
   n >M_1 \frac{d^2}{\alpha^2} \max\left\{e^{4k!\beta d},1\right\}\log \frac{\bpk}{\varepsilon},
\end{equation*}
then the following holds with probability at least $1-\varepsilon$:
\begin{equation*}
    \|\hat{\bm J}_{r,P}-\bm J_{r}\|_2\leq \frac{M_2\sqrt{d}e^{2k!\beta d}}{\alpha k!} \sqrt{\frac{\log\frac{\bpk}{\varepsilon}}{n}}.
\end{equation*}
\end{theorem}


\begin{proof}
Let $\gamma \in (0,1)$ which will be chosen suitably later. In view of Lemma \ref{lem8}, we know that with probability at least $1-\gamma\varepsilon$, Condition \ref{cond1} is satisfied with $$\lambda := 2k!\sqrt{\frac{8\log\left(2\bpk/\gamma\varepsilon\right)}{n}}~.$$ Next, note that for some $\eta_r \in K$, one has by the Cauchy-Schwarz inequality,
$$\|\eta_r\|_1\le 4\sqrt{d} \|\eta_r\|_2~.$$ Define $R:= c/\sqrt{d} k!$, where $c$ is a constant to be chosen later. Then, in view of Lemma \ref{lemsv}, Condition \ref{cond2} is satisfied with $$\kappa := \frac{\alpha (k!)^2 e^{-2k!\beta d}}{4(1+4c)}$$ with probability at least $1-(1-\gamma)\varepsilon$  as long as $n > 2^{11} \frac{d^2}{\alpha^2}\log\frac{2\bpk^2}{(1-\gamma)\varepsilon}$. Now, we can apply Proposition \ref{prop11} as long as $\frac{c}{\sqrt{d}k!}>3\frac{\sqrt{d} \lambda}{\kappa}$, which is equivalent to the condition:
$$n > \frac{C_1d^2}{\alpha^2} e^{4k!\beta d}\log\left(\frac{2\bpk}{\gamma\varepsilon}\right)$$
for some constant $C_1>0$. In that case, we have:
$$ \|\hat{\bm J}_{r,P}-\bm J_{r}\|\leq \frac{L\sqrt{d} e^{2k!\beta d}}{\alpha k!} \sqrt{\frac{\log \frac{2\bpk}{\gamma \varepsilon}}{n}}$$
for some constant $L>0$. Theorem \ref{thm1} now follows, by taking for example, $\gamma = 1/2$ and $c=1$.
\end{proof}



\subsection{The Regularized Interaction Screening Estimator}
In this section, we analyze the performance of the RISE. The main result about the rate of convergence of the RISE is stated below: 

\begin{theorem}\label{thm2}
Suppose that $d$ denotes the maximum degree of the hypernetwork with adjacency $\bm J$, and the regularization parameter $\lambda$ is chosen as:
\begin{equation*}\label{reg2}
    \lambda =  2\sqrt{2}k!e^{k!\beta d }\sqrt{\frac{\log\frac{4\bpk}{\varepsilon}}{n}},
\end{equation*}
Then, there exist constants $M_1,M_2>0$, such that for every node $r\in V$ and any $\varepsilon \in (0,1)$, if
\begin{equation*}
   n >M_1 \frac{d^2}{\alpha^2}\max\{e^{4k!\beta d }, 1\} \log\frac{\bpk}{\varepsilon},
\end{equation*}
the following properties hold with probability at least $1-\varepsilon$:
\begin{equation*}
    \|\hat{\bm J}_{r,I}-\bm J_{r}\|\leq \frac{M_2\sqrt{d} e^{2k!\beta d}}{\alpha k!} \sqrt{\frac{\log\frac{\bpk}{\varepsilon}}{n}}.
\end{equation*}
\end{theorem}

\begin{proof}
In view of Lemma \ref{lem8888}, we know that with probability at least $1-\frac{\varepsilon}{2}$, Condition \ref{cond1} is satisfied with
$$\lambda := 2\sqrt{2}k!e^{k!\beta d }\sqrt{\frac{\log\frac{4\bpk}{\varepsilon}}{n}}$$
If we set $R = \frac{2}{\sqrt{d} k!}$, then in view of Lemma \ref{lem77@}, Condition \ref{cond2} is satisfied with 
$$\kappa := \frac{\alpha (k!)^2 e^{-k!\beta d}}{20}$$
with probability at least $1-\frac{\varepsilon}{2}$ as long as $n> 2^{11} \frac{d^2}{\alpha^2} \log \frac{4\bpk^2}{\varepsilon}$. Now, we can apply Proposition \ref{prop11} as long as $\frac{2}{\sqrt{d} k!} > 3\frac{\sqrt{d}\lambda}{\kappa}$, which is equivalent to the condition:
$$n > \frac{Cd^2}{\alpha^2} e^{4k!\beta d} \log\frac{4\bpk}{\varepsilon}$$
for some constant $C>0$.  In that case, we have:
$$\|\hat{\bm J}_{r,I}-\bm J_{r}\|\leq \frac{D\sqrt{d} e^{2k!\beta d}}{\alpha k!} \sqrt{\frac{\log \frac{4\bpk}{\varepsilon}}{n}}$$ for some constant $L>0$, which completes the proof of Theorem \ref{thm2}. 
\end{proof}

\section{Simulation Studies}\label{sec:simul}
In this section, we present some numerical experiments that illustrate the comparative performance of the two tensor recovery algorithms. The Julia package \textit{GraphicalModelLearning} is modified and used to learn the Ising models with the interaction screening and pseudo-likelihood approaches. The performances
of the RPLE and RISE are dependent on the regularization coefficient $\lambda$, which was set as $\lambda := c\sqrt{\log(4\bpk / \varepsilon)/n}$, where $c$ is a constant tuned according to the Bayesian Information Criterion (BIC). To be precise, for each node $r$, the optimal value of $\lambda$ is tuned by minimizing the BIC value:
$$\mathrm{BIC}_r(\lambda) := \mathcal{L}_p(\bm \hat{J}_{r,\lambda};\mathfrak{X}^n) + \mathrm{df}(\lambda)\log p$$
over a grid of values of $\lambda$, where $\mathcal{L}$ denotes the objective function, which is $\mathcal{S}$ for the RISE and $\ell$ for the RPLE, and $\hat{J}_{r,\lambda}$ is the estimate corresponding to the penalty parameter $\lambda$.


\begin{figure}[h]
    \centering
    \begin{subfigure}[b]{0.48\textwidth}
        \centering
        \includegraphics[width=7cm]{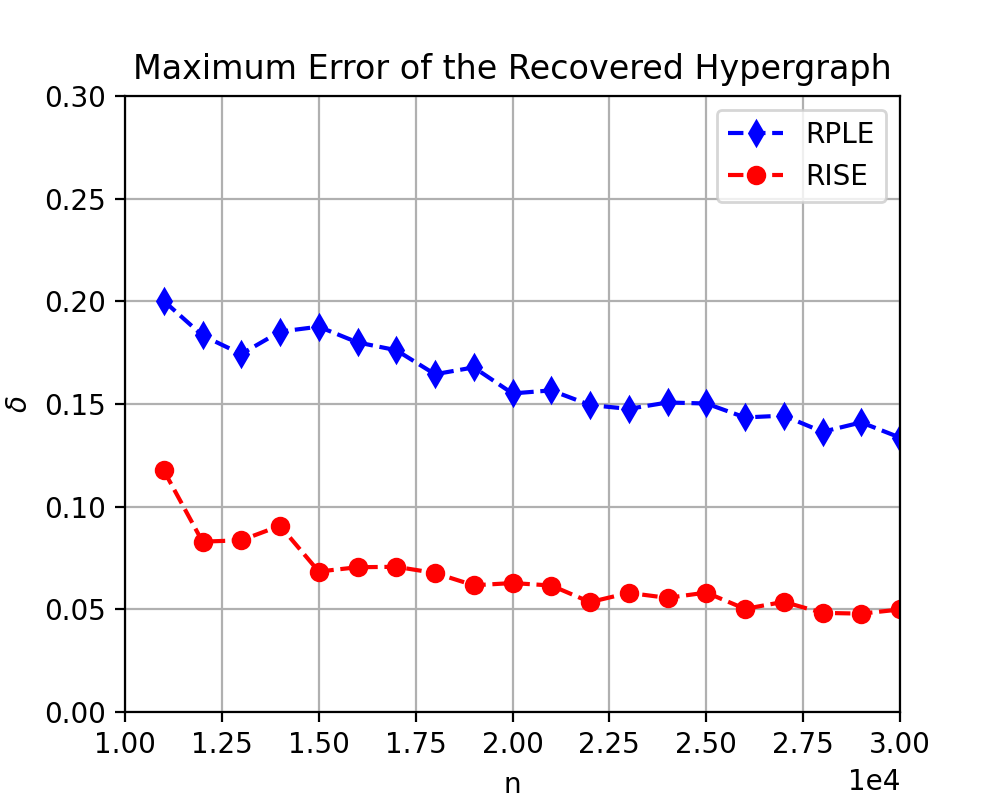}
        \caption[Network2]%
            {{\small $\beta = 1$}}    
    \end{subfigure}
    \hfill
    \begin{subfigure}[b]{0.48\textwidth}  
        \centering 
        \includegraphics[width=7cm]{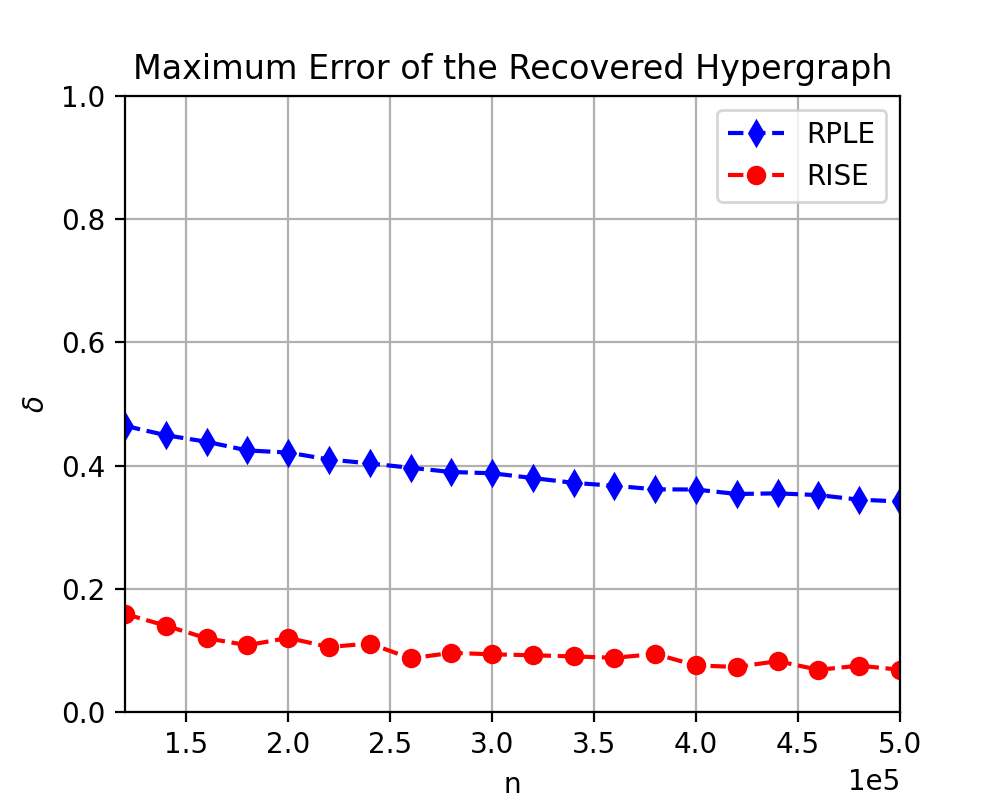}
        \caption[]%
            {{\small $\beta = 1.5$}}    
    \end{subfigure}
    \vskip\baselineskip
        \begin{subfigure}[b]{0.48\textwidth}   
            \centering 
            \includegraphics[width=7cm]{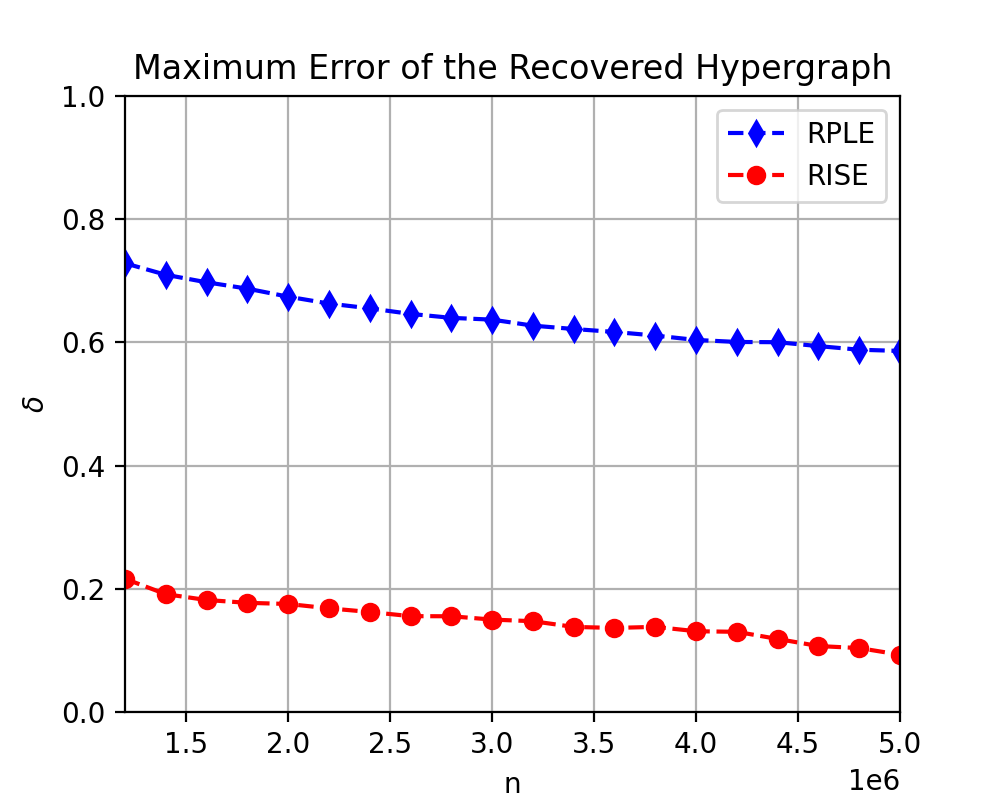}
            \caption[]%
            {{\small $\beta = 2$}}    
        \end{subfigure}
        \hfill
        \begin{subfigure}[b]{0.48\textwidth}   
            \centering 
            \includegraphics[width=7cm]{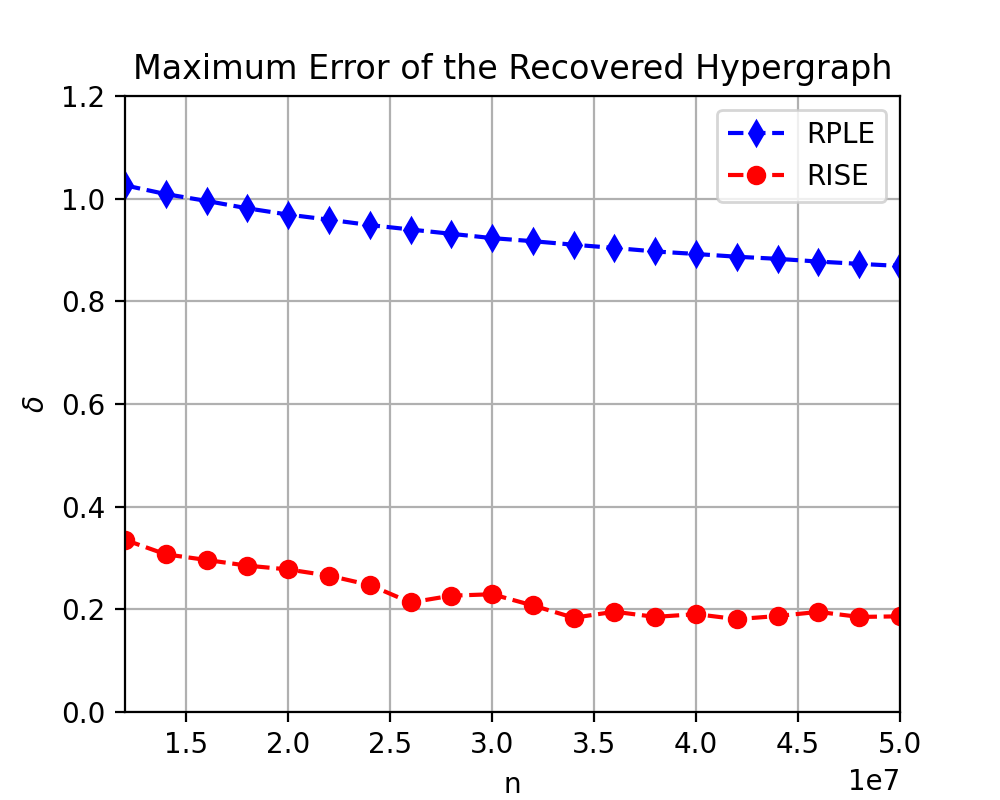}
            \caption[]%
            {{\small $\beta = 2.5$}}    
        \end{subfigure}
        \caption[maximum error ]
        {\small Plot of the estimation error (vertical axis) against the sample size $n$ (horizontal axis), for a $3$-uniform, $3$-regular hypergraph on $16$ nodes} 
        \label{fig_ee}
\end{figure}

\begin{figure}[h]
    \centering
    \includegraphics[width=0.5\textwidth]{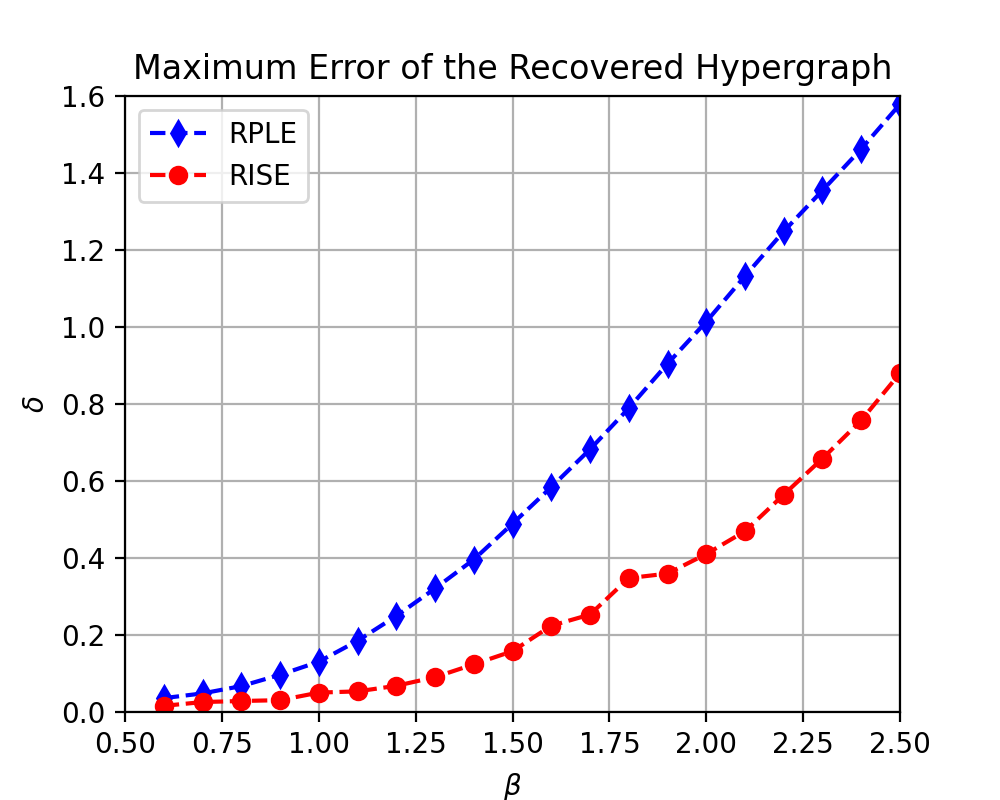}
    \caption{Plot of the estimation error (vertical axis) against the maximum coupling intensity $\beta$ (horizontal axis) for sample size $n = 10^5$}
    \label{fig5}
\end{figure}

For our numerical experiments, we work with the Ising model on $3-$regular, $3-$uniform hypergraphs on $16$ nodes. In Figure \ref{fig_ee}, we plot the estimation error of the recovered hypergraphs against the sample size $n$, for four values of the coupling intensity $\beta$, namely $1, 1.5, 2, 2.5$. We observe that the estimation errors are higher for both the RPLE and the RISE for higher values of $\beta$, as expected. Further, the simulations suggest that the RISE performs better than the RPLE in terms of the estimation error, which leaves open the natural question of whether this better performance of the RISE can be demonstrated theoretically. In Figure \ref{fig5}, we plot the estimation error for a fixed sample size $n = 10^5$ against varying coupling strengths $\beta$. The figure clearly shows an exponentially increasing dependence of the estimation error on $\beta$, as suggested in Theorems \ref{thm1} and \ref{thm2}.



\section{Real Data Analysis}\label{sec:realdata}
In this section, we implement our methods on gene data from the Curated Microarray Database (\href{https://sbcb.inf.ufrgs.br/cumida}{CuMiDa})\cite{sbcb}, a database composed of 78 handpicked microarray data sets for Homo sapiens that were carefully examined from more than 30,000 microarray experiments from the Gene Expression Omnibus using a rigorous filtering criterion. All data sets were individually submitted to background correction, normalization, sample quality analysis, and were manually edited to eliminate erroneous probes.

The data from the microarray experiment consists of a two-dimensional matrix with genes as rows and samples as columns (coming from different conditions). Each cell in the matrix is a normalized real number indicating how much a gene is expressed in a sample. These expression matrices will usually have thousands of rows and dozens or hundreds of columns. We selected the one particular data set with the largest sample size to facilitate our analysis. The $Liver\_GSE14520\_U133A$ data set contains 357 samples, of which 181 samples have Hepatocellular carcinoma (HCC), and 176 samples are normal. For each sample there are 22279 genes. To reduce the dimension of the data, feature selection was conducted through the Random Forest algorithm. A set of 51 genes was selected as the nodes in the tensor Ising model. The correspondence between the genes and node numbers is given in Appendix \ref{sec:genes}. The values for each node were standardized to binary values $\{-1,+1\}$ by mapping original gene expression values above the mean to $+1$ and values below the mean to $-1$. We then fitted the data of HCC samples and normal samples to a 3-spin Ising model separately, and recovered the corresponding tensors associated to the HCC patients and normal people, using the RISE algorithm.

For both the normal and HCC samples, the fitted hyperedges were then ordered by the absolute value of their coefficients, and the vertices were ordered according to their degrees in the tables above. In Table \ref{tab1}, it shows that the group effects of genes X200910\_at, X212551\_at and X217165\_x\_at, and the genes X203554\_x\_at, X212551\_at and X217165\_x\_at have highest weights in normal samples. From Table \ref{tab2}, we notice that the group effect of genes X210481\_s\_at, X216025\_x\_at and X216661\_x\_at, and genes X201293\_x\_at, X210481\_s\_at and X212661\_x\_at have higher weights in HCC samples. Tables \ref{tab3} and \ref{tab4} illustrate that the genes X201453\_x\_at and X217165\_x\_at are more important in the samples of normal people, and genes X210481\_s\_at and X207609\_s\_at are more related to Hepatocellular carcinoma.

\begin{table}[t!]
\captionsetup{font=small}
\caption{Triangles recovered from normal samples ordered by absolute value of edge weights}
\label{tab1}
\centering
\vskip-.3cm
\hrule
\smallskip
\scriptsize
\begin{tabular}{ccccccccc}
(3,33,40) & (10,33,40) & (13,28,39)
& (13,28,46) & (13,28,38) & (6,9,46)
& (6,9,39) & (6,9,38) & (40,41,48)
\\
0.3375&0.3375
&0.2986&0.2986
&0.2986&0.2874
&0.2874&0.2874&0.2723
\\
\end{tabular}
\hrule
\end{table}

\begin{table}[t!]
\captionsetup{font=small}
\caption{Triangles recovered from HCC samples ordered by absolute value of edge weights}
\label{tab2}
\centering
\vskip-.3cm
\hrule
\smallskip
\scriptsize
\begin{tabular}{ccccccccc}
(29,38,39)&
(5,29,35)&
(16,29,50)&
(4,10,23)&
(14,18,24)&
(29,36,40)&
(21,29,42)&
(14,47,49)&
(29,31,35)
\\
-0.5683&
-0.5363&
-0.4622&
-0.4358&
-0.4352&
-0.4144&
-0.4041&
-0.3871&
-0.3853
\\
\end{tabular}
\hrule
\end{table}

\begin{table}[t!]
\captionsetup{font=small}
\caption{Frequency of nodes in hyperedges recovered from normal samples}
\label{tab3}
\centering
\vskip-.3cm
\hrule
\smallskip
\scriptsize
\begin{tabular}{c|ccccccccc}
Node &
6&
40&
48&
33&
9&
13&
28&
10&
3

\\
Frequency&
11&
6&
4&
4&
3&
3&
3&
3&
3
\\
\end{tabular}
\hrule
\end{table}

\begin{table}[t!]
\captionsetup{font=small}
\caption{Frequency of nodes in hyperedges recovered from HCC samples}
\label{tab4}
\centering
\vskip-.3cm
\hrule
\smallskip
\scriptsize
\begin{tabular}{c|cccccccc}
Node &
29&
23&
14&
28&
10&
47&
16&
50

\\
Frequency&
13&
5&
4&
4&
3&
3&
3&
3

\\
\end{tabular}
\hrule
\end{table}

\appendix
\section{Verification of Conditions \ref{cond1} and \ref{cond2} for the Pseudolikelihood}\label{lem133apx}
In this section, we verify conditions \ref{cond1} and \ref{cond2} for the pseudolikelihood, which is a crucial step in the proof of Theorem \ref{thm1}. We begin by verifying Condition \ref{cond1}.

\begin{lem}\label{lem8}
  For any $\varepsilon > 0$, with probability at least $1-\varepsilon$, we have:
    \begin{equation*}
    \left\|\nabla \ell(\bm J_r;\mathfrak{X}^n) \right\|_{\infty} \leq k! \sqrt{\frac{8\log\left(2\bpk/\varepsilon\right)}{n}},
    \end{equation*}
\end{lem}
\begin{proof}
To begin with, note that:
$$\frac{\partial \ell(\bm J_r;\mathfrak{X}^n)}{\partial J_{r,r_1,\ldots,r_{k-1}}} = -\frac{1}{n}\sum_{i=1}^n k! X_{r_1}^{(i)} \ldots X_{r_{k-1}}^{(i)} \left(X_r^{(i)}-\tanh\left(k m_r(\bm X^{(i)})\right)\right) =: -\frac{1}{n}\sum_{i=1}^n Y_i$$
where $Y_1,\ldots,Y_n$ are mean $0$, independent random variables bounded between $-2k!$ and $2k!$. Hence, by Hoeffding's inequality, we have:
$$\p\left(\left|\frac{\partial \ell(\bm J_r;\mathfrak{X}^n)}{\partial J_{r,r_1,\ldots,r_{k-1}}}\right| \ge \frac{\textcolor{black}{2\sqrt{2}} k!t}{\sqrt{n}}\right) \le 2e^{-t^2}~.$$ Choosing $t := \sqrt{\log\left(2\bpk/\varepsilon\right)}$, we thus have:
$$\p\left(\left|\frac{\partial \ell(\bm J_r;\mathfrak{X}^n)}{\partial J_{r,r_1,\ldots,r_{k-1}}}\right| \ge k!\sqrt{\frac{8\log\left(2\bpk/\varepsilon\right)}{n}}\right) \le \frac{\varepsilon}{\bpk}~.$$
It now follows from a simple union bound, that:
$$\p\left(\left\|\nabla \ell(\bm J_r;\mathfrak{X}^n) \right\|_{\infty} \ge k!\sqrt{\frac{8\log\left(2\bpk/\varepsilon\right)}{n}}\right) \le \varepsilon~.$$
The proof of Lemma \ref{lem8} is now complete.
\end{proof}

 Next, we verify Condition \ref{cond2}. Towards this, first note that  for any fixed node $r\in [p]$, the Hessian matrix of the negative log pseudolikelihood is given by:
\begin{equation*}
    \nabla^2 \ell(\bm J_r;\mathfrak{X}^n)= -\frac{1}{n} \sum_{i=1}^n \delta_r(\bm X^{(i)};\bm J) \bm X^{(i)}_{\cdot r} (\bm X^{(i)}_{\cdot r})^\top,
\end{equation*}
where
\begin{equation*}
   \delta_r(\bm X;\bm J):=\frac{4(k!)^2 e^{2k X_r m_r(\bm X)}}{(e^{2k X_r m_r(\bm X)}+1)^2}.
\end{equation*}
and recall that the $\binom{p-1}{k-1}$ dimensional vector $\bm X_{\cdot r}$ is defined as 
\begin{equation*}
    \bm X_{\cdot r}:=(X_{r_1}\cdots X_{r_{k-1}})_{(r_1,\ldots,r_{k-1})\in T_r}~.
\end{equation*}
Next, for $\eta_r \in \mathbb{R}^\bpk$, define the remainder of the first order Taylor expansion of $\ell(\bm J_r + \eta_r;\mathfrak{X}^n)$ as:
\begin{equation*}
        \Delta(\bm J_r;\mathfrak{X}^n ):=\ell(\bm J_r + \eta_r;\mathfrak{X}^n) - \ell(\bm J_r;\mathfrak{X}^n) - \langle \nabla \ell(\bm J_r;\mathfrak{X}^n), \eta_r\rangle.
\end{equation*}
In the next lemma, we lower bound this remainder term.
\begin{lem}\label{rembnd}
    For all $\eta_r \in \mathbb R^{\bpk}$, the remainder of the first order Taylor expansion is bounded below by:
    \begin{equation*}
        \Delta(\bm J_r;\mathfrak{X}^n ) \geq \frac{e^{-2k!\beta d}(k!)^2 }{2 k! \|\eta_r\|_1+2} ~\eta_r^\top \hat{Q} \eta_r
    \end{equation*}
    where $\hat{Q}:= \frac{1}{n} \sum_{i=1}^n \bm X^{(i)}_{\cdot r} (\bm X^{(i)}_{\cdot r})^\top$.
\end{lem}
\begin{proof}
   To begin with, note that:
    \begin{equation*}
        \Delta(\bm J_r;\mathfrak{X}^n) = \frac{1}{n}\sum_{i=1}^n \left [ \log \left(\frac{\cosh(km_r(\bm X^{(i)}) + k!\eta_r^\top \bm X_{\cdot r}^{(i)}  )}{\cosh(km_r(\bm X^{(i)}))}\right)-k! \eta_r^\top \bm X_{\cdot r}^{(i)} \tanh(km_r(\bm X^{(i)})) \right]
    \end{equation*}
Using Lemma \ref{logcsh1} with $x := km_r(\bm X^{(i)})$ and $a:= k!\eta_r^\top \bm X_{\cdot r}^{(i)}$, and the inequality $$1-\tanh^2(x) \ge e^{-2|x|}~,$$ we have:
$$\Delta(\bm J_r;\mathfrak{X}^n) \ge \frac{\exp\left(-2\left|km_r(\bm X^{(i)})\right|\right) (k!)^2}{2+2k!\|\eta_r\|_1} ~\eta_r^\top \hat{Q} \eta_r~.$$
    Lemma \ref{rembnd} now follows on observing that $|km_r(\bm X)|\leq k!\beta d$.
\end{proof}

In the following lemma, we provide error bounds for estimating $Q$ by $\hat{Q}$.

\begin{lem}\label{qaprx}
    For every $\delta > 0$, $\varepsilon > 0$ and $n\ge \frac{2}{\delta^2}\log \frac{2\binom{p-1}{k-1}^2}{\varepsilon}$, we have the following with probability at least $1-\varepsilon$, we have the following: 
    \begin{equation*}
        \max_{(t_r,t_s)\in T_r^2}|\hat{Q}_{t_r, t_s} - Q_{t_r, t_s}|< \delta,
    \end{equation*}
\end{lem}
\paragraph{Proof}
Note that for $t_r:=(r_1,\ldots,r_{k-1})$ and $t_s:=(s_1,\ldots,s_k)$, we have
\begin{equation*}
    \hat{Q}_{t_r, t_s} = \frac{1}{n} \sum_{i=1}^n x_{r_1}...x_{r_{k-1}}x_{s_1}...x_{s_{k-1}},
\end{equation*}
Since $|x_{r_1}...x_{r_{k-1}}x_{s_1}...x_{s_{k-1}}|\leq 1$, by Hoeffding's inequality we have
\begin{equation*}
    P\left(|\hat{Q}_{t_r t_s} - Q_{t_r t_s}|\geq \delta\right)\leq 2\exp\left(-\frac{n\delta^2}{2}\right),
\end{equation*}
Using a union bound over the upper triangle of the symmetric matrix we have
\begin{equation*}
    P\left( \max_{(t_r,t_s)\in T_r^2} |\hat{Q}_{t_r t_s} - Q_{t_r t_s}|< \delta\right) \geq 1 - 2\bpk ^2 \exp\left(-\frac{n\delta^2}{2}\right) \ge 1-\varepsilon
\end{equation*}
thereby completing the proof of Lemma \ref{qaprx}.

The following lemma provides the final step towards verifying Condition \ref{cond2} for the pseudolikelihood:

\begin{lem}\label{lemsv}
    For all $\varepsilon> 0$, $R > 0$, and $n> 2^{11}\frac{d^2}{\alpha^2}\log \frac{2\bpk ^2}{\varepsilon}$, with probability at least $1-\varepsilon$, the restricted strong convexity condition holds, i.e.
    \begin{equation*}
        \Delta(\bm J_r;\mathfrak{X}^n ) \geq \frac{\alpha (k!)^2 e^{-2k!\beta d}}{4(1+4k!\sqrt{d}R)}\|\eta_r\|_2^2,
    \end{equation*}
   for all $\eta_r \in K$ such that $\|\eta_r\|_2 \leq R$.
\end{lem}
\paragraph{Proof}
From Lemma \ref{rembnd}, we have:
\begin{equation*}
        \Delta(\bm J_r;\mathfrak{X}^n ) \geq \frac{e^{-2k!\beta d}}{2 k!\|\eta_r\|_1+2} (k!)^2 \eta_r^\top \hat{Q} \eta_r.
    \end{equation*}
Since $\|\eta_r\|_1 \leq 4\sqrt{d} \|\eta_r\|_2\leq 4\sqrt{d} R$, we conclude that:
\begin{equation*}
    \Delta(\bm J_r;\mathfrak{X}^n ) \geq \frac{e^{-2k!\beta d}}{8 k!\sqrt{d}R+2} (k!)^2 \eta_r^\top \hat{Q} \eta_r,
\end{equation*}
Next, we have by Assumption \ref{asm1}: 
\begin{equation*}
    \begin{aligned}
    \eta_r^\top \hat{Q} \eta_r &=   \eta_r^\top Q \eta_r + \eta_r^\top (\hat{Q} -Q) \eta_r \\
    &\geq  \alpha\|\eta_r\|_2^2 + \eta_r^\top (\hat{Q} -Q) \eta_r,
    \end{aligned}
\end{equation*}
Using Lemma \ref{qaprx} with $\delta := \frac{\alpha}{32d}$, we have the following for $n>2^{11}\frac{d^2}{\alpha^2}\log \frac{2\bpk ^2}{\varepsilon}$ with probability at least $1-\varepsilon$, 
\begin{equation*}
    \begin{aligned}
        \eta_r^\top (\hat{Q} -Q) \eta_r &\geq - \frac{\alpha}{32d} \|\eta_r\|_1^2 \\
        & \geq - \frac{\alpha}{2} \|\eta_r\|_2^2,
    \end{aligned}
\end{equation*}
Hence, we have: 
\begin{equation*}
        \Delta(\bm J_r;\mathfrak{X}^n ) \geq \frac{\alpha (k!)^2 e^{-2k!\beta d}}{4(1+4 k!\sqrt{d}R)}\|\eta_r\|_2^2
\end{equation*}
which completes the proof of Lemma \ref{lemsv}.

\section{Verification of Conditions \ref{cond1} and \ref{cond2} for the Interaction Screening Approach}\label{lem144apx}

In this section, we verify conditions \ref{cond1} and \ref{cond2} for the interaction screening function, which is a fundamental step for proving Theorem \ref{thm2}. We start by checking Condition \ref{cond1}.

\begin{lem}\label{lem8888}
For any $\varepsilon > 0$, with probability at least $1-\varepsilon$, we have:
    \begin{equation*}
    \left\|\nabla \eS(\bm J_r;\mathfrak{X}^n) \right\|_{\infty} \leq \sqrt{2}k!e^{k!\beta d }\sqrt{\frac{\log\frac{2\bpk}{\epsilon}}{n}},
    \end{equation*}
\end{lem}
\begin{proof}
First, note that we have:
$$\frac{\partial \eS(\bm J_r;x^{(i)})}{\partial J_{r,r_1,\ldots,r_{k-1}}} = -\frac{1}{n}\sum_{i=1}^n k! X_{r_1}^{(i)} \ldots X_{r_{k-1}}^{(i)} X_r^{(i)} e^{-k X_r^{(i)}m_r(\bm X^{(i)})} := -\frac{1}{n}\sum_{i=1}^n V_i$$ where
\begin{equation*}
    |V_i|\leq k!e^{k!\beta d},
\end{equation*}
Since $V_1,\ldots,V_n$ are i.i.d. with mean $0$, we have the following by Hoeffding’s inequality:
\begin{equation*}
    P\left(\left|\frac{\partial \eS(\bm J_r;\mathfrak{X}^n) }{\partial \bm J_{r,r_1,...,r_{k-1}}}\right|\geq \frac{\sqrt{2} k!e^{k!\beta d} t}{\sqrt{n}}\right)\leq 2 e^{-t^2},
\end{equation*}
As there are $\bpk$ components of the gradient vector, we can choose $t := \sqrt{\log\left(2\bpk/\epsilon\right)}$. Using a union bound we get the following with probability at least $1-\epsilon$:
\begin{equation*}
    \left\|\nabla \eS(\bm J_r;\mathfrak{X}^n) \right\|_{\infty} \leq \sqrt{2}k!e^{k!\beta d }\sqrt{\frac{\log\frac{2\bpk}{\epsilon}}{n}}~,
\end{equation*}    
thereby completing the proof of Lemma \ref{lem8888}.
\end{proof}

Next we check condition 2. Towards this, first note that for any fixed node $r\in V$, we can express the Hessian of the interaction screening function as:
\begin{equation*}
    \nabla^2 \eS(\bm J_r;\mathfrak{X}^n)=\frac{1}{n} \sum_{i=1}^n \gamma_r(\bm X^{(i)};\bm J_r) \bm X^{(i)}_{\cdot r} (\bm X^{(i)}_{\cdot r})^\top,
\end{equation*}
where
\begin{equation*}
   \gamma_r(\bm X;\bm J_r):= (k!)^2\exp (- k X_r m_r(\bm X)),
\end{equation*}
Using the inequality $\gamma_r(\bm X;\bm J) \geq (k!)^2 e^{-\beta d (k-1)!}$, the Hessian is lower bounded in the positive semi-definite sense by:
\begin{equation*}
    \nabla^2 \eS(\bm J_r;\mathfrak{X}^n) \succcurlyeq (k!)^2 e^{-\beta d k!} \hat{Q}
\end{equation*}
where recall that $\hat{Q} = \frac{1}{n} \sum_{i=1}^n \bm X^{(i)}_{\cdot r} (\bm X^{(i)}_{\cdot r})^\top$.

As before, for every $\eta_r \in \mathbb{R}^\bpk$, define the remainder term:
\begin{equation*}
        \Delta(\bm J_r;\mathfrak{X}^n ):=\eS(\bm J_r + \eta_r;\mathfrak{X}^n) - \eS(\bm J_r;\mathfrak{X}^n) - \langle \nabla \eS(\bm J_r;\mathfrak{X}^n), \eta_r\rangle.
\end{equation*}
In the next lemma, we lower bound this remainder term.
\begin{lem}
    For all $\eta_r \in \mathbb R^{\bpk}$, the remainder of the first order Taylor expansion is bounded below by:
    \begin{equation*}
        \Delta(\bm J_r;\mathfrak{X}^n ) \geq \frac{(k!)^2 e^{-k!\beta d}}{2 + k!\|\eta_r\|_1} \eta_r^\top \hat{Q} \eta_r.
    \end{equation*}
\end{lem}
\begin{proof}
To begin with, note that:
    \begin{equation*}
        \Delta(\bm J_r;\mathfrak{X}^n ) = \frac{1}{n}\sum_{i=1}^n \left [ e^{-kX_r^{(i)}m_r(\bm X^{(i)})}\left( e^{-k!\eta_r^\top \bm X_{\cdot r}^{(i)}X_r^{(i)}}-1+k!\eta_r^\top X_r^{(i)} \bm X_{\cdot r}^{(i)}\right) \right],
    \end{equation*}
Since $|km_r(\bm X)|\leq k!\beta d$, we have:
\begin{equation*}
    e^{-kX_r^{(i)}m_r(\bm X^{(i)})} \geq e^{-k!\beta d},
\end{equation*}
Next, using Lemma \ref{expineq7} together with the inequality 
\begin{equation*}
    \left |k!\eta_r^\top X_r^{(i)} \bm X_{\cdot r}^{(i)}\right| \leq k!\| \eta_r\|_1,
\end{equation*}
we have 
\begin{equation*}
\begin{aligned}
    \Delta(\bm J_r;\mathfrak{X}^n ) & \geq \frac{e^{-k!\beta d}}{2 + k!\|\eta_r\|_1}\frac{1}{n}\sum_{i=1}^n \left( k!X_r^{(i)}\eta_r^\top \bm X_{\cdot r}^{(i)} \right)^2\\
    & = \frac{(k!)^2 e^{-k!\beta d}}{2 + k!\|\eta_r\|_1} \eta_r^\top \hat{Q} \eta_r ,
\end{aligned}
\end{equation*}

\end{proof}

The following lemma provides the final step towards verifying Condition \ref{cond2} for
the interaction screening approach:

\begin{lem}\label{lem77@}
 For all $\varepsilon> 0$, $R > 0$, and $n> 2^{11}\frac{d^2}{\alpha^2}\log \frac{2\bpk ^2}{\varepsilon}$, with probability at least $1-\varepsilon$, the restricted strong convexity condition holds, ie
    \begin{equation*}
        \Delta(\bm J_r;\mathfrak{X}^n ) \geq \frac{\alpha (k!)^2 e^{-k!\beta d}}{4(1+2 k!\sqrt{d}R)}\|\eta_r\|_2^2,
    \end{equation*}
    for all $\eta_r \in K$ such that $\|\eta_r\|_2 \leq R$.
\end{lem}
\begin{proof}
From Lemma 6, we have
\begin{equation*}
        \Delta(\bm J_r;\mathfrak{X}^n ) \geq \frac{(k!)^2 e^{-k!\beta d}}{2 + k!\|\eta_r\|_1} \eta_r^\top \hat{Q} \eta_r.
    \end{equation*}
Since $\|\eta_r\|_1 \leq 4\sqrt{d} \|\eta_r\|_2\leq 4\sqrt{d} R$, we have:
\begin{equation*}
    \Delta(\bm J_r;\mathfrak{X}^n ) \geq \frac{(k!)^2 e^{-k!\beta d}}{4 k!\sqrt{d}R+2}  \eta_r^\top \hat{Q} \eta_r,
\end{equation*}
Next, we have by Assumption \ref{asm1}:
\begin{equation*}
    \begin{aligned}
    \eta_r^\top \hat{Q} \eta_r &=   \eta_r^\top Q \eta_r + \eta_r^\top (\hat{Q} -Q) \eta_r \\
    &\geq  \alpha\|\eta_r\|_2^2 + \eta_r^\top (\hat{Q} -Q) \eta_r,
    \end{aligned}
\end{equation*}
Using Lemma 3, we set $\delta = \frac{\alpha}{32d}$, we get for $n>2^{11}\frac{d^2}{\alpha^2}\log \frac{2\bpk ^2}{\varepsilon}$ with probability at least $1-\varepsilon$, 
\begin{equation*}
    \begin{aligned}
        \eta_r^\top (\hat{Q} -Q) \eta_r &\geq - \frac{\alpha}{32d} \|\eta_r\|_1^2 \\
        & \geq - \frac{\alpha}{2} \|\eta_r\|_2^2,
    \end{aligned}
\end{equation*}
Thus 
\begin{equation*}
        \Delta(\bm J_r;\mathfrak{X}^n ) \geq \frac{\alpha (k!)^2 e^{-k!\beta d}}{4(1+2 k!\sqrt{d}R)}\|\eta_r\|_2^2.
\end{equation*}
This completes the proof of Lemma \ref{lem77@}.
\end{proof}

\section{Technical Lemmas}\label{sec:techlem}
In this section, we prove some technical lemmas necessary for the proofs of the main results.
\begin{lem}\label{logcsh1}
    For every $x,a\in \mathbb{R}$, we have:
    \begin{equation*}
        \log\left(\frac{\cosh(x+a)}{\cosh(x)}\right)-a \tanh(x) \geq \frac{(1-\tanh^2(x))a^2}{2+2|a|},
    \end{equation*}
\end{lem}
\begin{proof}
  Fixing $x\in \mathbb{R}$, let
\begin{equation*}
    f(a) := \log\left(\frac{\cosh(x+a)}{\cosh(x)}\right)-a \tanh(x) - \frac{\sech^2(x) a^2}{2+2|a|},
\end{equation*}
First, consider the case $a>0$, and note that:
\begin{equation*}
    f'(a)= \tanh(x+a)-\tanh(x)-\frac{1}{2}\sech^2(x)\left(1-\frac{1}{(a+1)^2}\right),
\end{equation*}
and
\begin{equation*}
    f''(a) = \sech^2(x+a)-\sech^2(x)\cdot \frac{1}{(a+1)^3},
\end{equation*}

We claim that there exists $\delta>0$ such that $f''>0$ on $(0,\delta)$ and $f''<0$ on $(\delta,\infty)$. Given this claim, we know that since $f'(0) = 0$ and $\lim_{a\rightarrow \infty} f'(a) = \frac{1}{2}(\tanh(x)-1)^2>0$, we must have $f'(a)>0$ for all $a>0$. Since $f(0)=0$, we must thus have $f(a)>0$ for all $a>0$, which proves Lemma \ref{logcsh1} for all $x\in \mathbb{R}$ and $a\ge 0$. If $a<0$, then note that we can apply this result with $x$ replaced by $-x$ and $a$ replaced by $-a$, to complete the proof of Lemma \ref{logcsh1}. Hence, it only suffices to prove the claim, in order to complete the proof of Lemma \ref{logcsh1}. Towards this, define $g(a) :=\cosh^2(x)\cdot (a+1)^3 - \cosh^2 (x+a)$,
whence
\begin{equation*}
\begin{aligned}
    g'(a) & = 3\cosh^2(x)(a+1)^2-\sinh(2x+2a)\\
    g''(a) & = 6 \cosh^2(x)(a+1)-2\cosh(2x+2a)\\
    g^{(3)}(a) & = 6\cosh^2(x) - 4\sinh(2x+2a),
\end{aligned}
\end{equation*}

Note that $g^{(3)}$ is strictly decreasing and $\lim_{a\rightarrow \infty}g^{(3)}(a) = -\infty$. Meanwhile, $g''(0) = 2\cosh^2(x)+2 >0$ and $\lim_{a\rightarrow \infty} g''(a) = -\infty$. Hence, $g''$ is positive on $(0,K)$ and negative on $(K,\infty)$ for some $K>0$.
Next, since $g'(0)=\cosh(x)(2e^{-x}+\cosh(x))>0$ and $\lim_{a\rightarrow \infty} g'(a) = -\infty$, the previous argument implies that $g'$ is positive on $(0,L)$ and negative on $(L,\infty)$ for some $L>0$.
Finally, since $g(0)=0$ and $\lim_{a\rightarrow \infty} g(a) = -\infty$, there exists $\delta>0$, such that for $0<a<\delta$, $\cosh^2(x)\cdot (a+1)^3 > \cosh^2 (x+a)$, and for $a>\delta$, $\cosh^2(x)\cdot (a+1)^3 < \cosh^2 (x+a)$. So for $0<a<\delta$, $f''(a)>0$, and for $a>\delta$, $f''(a)<0$. This proves the claim, and completes the proof of Lemma \ref{logcsh1}.
\end{proof}

\begin{lem}\label{expineq7}
    For all $x\in \mathbb{R}$, we have:
    $$e^{-x} -1+x \ge \frac{x^2}{2+|x|}\quad\text{for all}~x\in \mathbb{R}$$
\end{lem}
\begin{proof}
    Define:
    $$g(x) = e^{-x}-1+x-\frac{x^2}{2+|x|}~.$$
    Then, for $x>0$, we have: $$g'(x) = -e^{-x} + \frac{4}{(2+x)^{2}} = \left(1+\frac{x}{2}\right)^{-2} - (e^{\frac{x}{2}})^{-2} >0~.$$ Since $g(0) = 0$, we must thus have $g(x) > 0$ for all $x>0$, thereby proving Lemma \ref{expineq7} for $x>0$. Now, suppose that $x<0$. Since $\sinh (-x) > -x$, we have:
    $$e^{-x}-1+x>e^x-1-x\ge \frac{x^2}{2+|x|}$$
    where the last inequality follows on applying Lemma \ref{expineq7} on $-x>0$.
\end{proof}

\section{Correspondence of node numbers and genes}\label{sec:genes}
In this section, we provide the correspondence between the node labels and the corresponding gene labels in the following tabular form. 

\begin{table}[h]
    \footnotesize
        \begin{tabular}[h]{c|c}
            \hline
            Node &Gene \\
            \hline
            1& X200734\_s\_at\\
            2& X200882\_s\_at\\
            3&	X200910\_at\\
            4&	X201128\_s\_at\\
            5&	X201293\_x\_at\\
            6&	X201453\_x\_at\\
            7&	X201656\_at\\
            8&	X202861\_at\\
            9&	X203316\_s\_at\\
            10&	X203554\_x\_at\\
            11&	X204428\_s\_at\\
            12&	X204641\_at\\
            13&	X205019\_s\_at\\
            14&	X205440\_s\_at\\
            15&	X205554\_s\_at\\
            16&	X205866\_at\\
            17&	X205890\_s\_at\\  \hline
        \end{tabular}
        \hfill
        \begin{tabular}[h]{c|c}
            \hline
            Node &Gene \\
            \hline
            18&	X205911\_at\\
            19&	X206380\_s\_at\\
            20&	X206453\_s\_at\\
            21&	X206643\_at\\
            22&	X206680\_at\\
            23&	X207609\_s\_at\\
            24&	X207804\_s\_at\\
            25&X207828\_s\_at\\
            26&	X207995\_s\_at\\
            27&	X209219\_at\\
            28& X209365\_s\_at\\
            29&	X210481\_s\_at\\
            30&	X211745\_x\_at\\
            31&	X211978\_x\_at\\
            32&	X212149\_at\\
            33&	X212551\_at\\
            34&	X212554\_at\\ \hline   
        \end{tabular}
        \hfill
        \begin{tabular}[h]{c|c}
            \hline
            Node &Gene \\
            \hline
            35&	X212661\_x\_at\\
            36&	X213629\_x\_at\\
            37&	X215138\_s\_at\\
            38&	X216025\_x\_at\\
            39&	X216661\_x\_at\\
            40&	X217165\_x\_at\\
            41&	X217319\_x\_at\\
            42&	X217521\_at\\
            43&	X218002\_s\_at\\
            44&	X218061\_at\\
            45&	X219787\_s\_at\\
            46&	X220017\_x\_at\\
            47&	X220114\_s\_at\\
            48&	X220148\_at\\
            49&	X220491\_at\\
            50&	X220496\_at\\
            51&	X36829\_at\\ \hline   
        \end{tabular}
    \caption{Correspondence of node and gene labels.}
    \label{tab:my_label}
\end{table}


\begin{thebibliography}{99}
\bibitem{nega}
Negahban, S. N., Ravikumar, P., Wainwright, M. and Yu, B. (2012), A Unified Framework for High-Dimensional Analysis of M-Estimators with Decomposable Regularizers. Statistical Science, 27(4).



\bibitem{structure_learning}
Anandkumar, A., Tan, V.Y.F., Huang, F, and Willsky, A.S. (2012), \textit{High-dimensional structure estimation in Ising models: Local separation criterion}, The Annals of Statistics, Vol. 40 (3), 1346--1375.

\bibitem{spatial}
Banerjee, S., Carlin, B.P., and Gelfand, A.E. (2014),
\textit{Hierarchical modeling and analysis for spatial data},
Chapman and Hall/CRC.

\bibitem{barra} 
Barra, A. (2009), \textit{Notes on ferromagnetic $p$-spin and REM}, Mathematical Methods in the Applied Sciences, 32 (7): 783–797.


\bibitem{montanarig}
Bento, J., \& Montanari, A. (2009), Which graphical models are difficult to learn?  Advances in Neural Information Processing Systems 22.

\bibitem{besag_lattice}
Besag, J. (1974), \textit{Spatial interaction and the statistical analysis of lattice systems}, J. Roy. Stat. Soc. B, Vol. 36, 192--236.

\bibitem{besag_nl}
Besag, J. (1975), \textit{Statistical analysis of non-lattice data}, The Statistician, Vol. 24 (3), 179--195. 

\bibitem{BM16}
Bhattacharya, B. and Mukherjee, S. (2018), \textit{Inference in ising models}, Bernoulli, Vol. 24 (1), 493--525. 




\bibitem{bovier} 
Bovier, A., Kurkova, I., and L\"{o}we, M. (2002), \textit{Fluctuations of the Free Energy in the REM and the p-Spin
SK Models}, The Annals of Probability, Vol. 30 (7): 605--651.

\bibitem{bresler}
Bresler, G. (2015),  \textit{Efficiently learning Ising models on arbitrary graphs}, Proceedings Symposium on Theory of Computing (STOC), 771--782.

\bibitem{bresler1}
Bresler, G. (2014),  \textit{Structure learning of antiferromagnetic Ising models}, Advances in Neural Information Processing Systems 27 (NIPS 2014).

\bibitem{bresler2}
Bresler, G. (2020),  \textit{Learning a tree-structured Ising model in order to make predictions}, The Annals of Statistics, 713--737.

\bibitem{fiftyfour}
Cantador, I., Brusilovsky, P. and Kuflik, T. (2011), \textit{Second workshop on information heterogeneity and fusion in recommender systems}, Proceedings of the 5th ACM Conference on Recommender Systems, RecSys'11, 387--388.

\bibitem{high_tempferro}
Cao, Y., Neykov, M. and Liu, H. (2019), \textit{High Temperature Structure Detection in Ferromagnets}, arXiv:1809.08204.

\bibitem{chatterjee} 
Chatterjee, S. (2007), \textit{Estimation in spin glasses: A first step},  The Annals of Statistics, Vol. 35 (5), 1931--1946.

\bibitem{discrete_tree}
Chow, C. and Liu, C. (1968), \textit{Approximating discrete probability distributions with dependence trees}, IEEE Transactions on Information Theory, Vol. 14~(3), 462--467.



\bibitem{cd_testing}
 Daskalakis, C., Dikkala, N. and Kamath, G. (2019), \textit{Testing Ising models}, IEEE Transactions on Information Theory, Vol. 65 (11), 6829--6852.

\bibitem{cd_trees}
Daskalakis, C., Mossel, E., and Roch, S. (2011), \textit{Evolutionary trees and the Ising model on the Bethe lattice: A proof of Steel's conjecture}, Probability Theory and Related Fields, Vol. 149 (1), 149--189.


\bibitem{luc} 
Devroye, L., Mehrabian, A. and Reddad, T. (2020), \textit{The minimax learning rates of normal and Ising undirected graphical models},  Electronic J. Statist. 14 (1).

\bibitem{gardner} 
Gardner, E. (1985), \textit{Spin glasses with $p$-spin interactions}, Nuclear Physics B, 257: 747--765.

\bibitem{sbcb}
Feltes, B. C., Chandelier, E. B., Grisci, B. I., \& Dorn, M. (2019), \textit{Cumida: an extensively curated microarray database for benchmarking and testing of machine learning approaches in cancer research}, Journal of Computational Biology, 26(4), 376-386.

\bibitem{pg_sm}
Ghosal, P. and Mukherjee, S. (2020), \textit{Joint estimation of parameters in Ising model}, The Annals of Statistics, Vol. 48(2), 785--810. 


\bibitem{disease}
Green, P.J. and Sylvia, R. (2002),
\textit{Hidden markov models and disease mapping},
Journal of the American Statistical Association, 97:1055--1070.

\bibitem{geman_graffinge}
Geman, S. and Graffigne, C. (1986), \textit{Markov random field image models and their applications
to computer vision}, Proceedings of the International Congress of Mathematicians, 1496--1517.

\bibitem{graphical_models_algorithmic}
Hamilton, L., Koehler, F. and Moitra, A. (2017), \textit{Information theoretic properties of Markov Random Fields, and their algorithmic applications}, Advances in Neural Information Processing Systems (NIPS), 2463--2472.

\bibitem{neural}
Hopfield, J.J. (1982),
\textit{ Neural networks and physical systems with emergent collective
  computational abilities},
Proc. Natl. Acad. Sci. USA, 79:2554--2558.

\bibitem{ising}
Ising, E. (1925), \textit{Beitrag zur theorie der ferromagnetismus}, Zeitschrift f\"ur Physik, Vol. 31, 253--258.

\bibitem{elhj1} 
Ibarrondo, R., Sanz, M. and Or\'us, R. (2022), \textit{Forecasting Election Polls with Spin Systems},  SN COMPUT. SCI. 3, 44.


\bibitem{isinggenapl5} 
Lipowski, A. (2022), \textit{Ising Model: Recent Developments and Exotic Applications}, Entropy (Basel). 24 (12): 1834.


\bibitem{liu}
Liu, T., Mukherjee, S. and Biswas, R. (2023), \textit{Tensor Recovery in High-Dimensional Ising Models}, arXiv preprint, arXiv:2304.00530v1.

\bibitem{lokhov} 
Lokhov, A.Y., Vuffray, M., Misra, S. and Chertkov, M. (2018), \textit{Optimal structure and parameter learning
of Ising models},  Science Advances, Vol. 4, Issue 3.

\bibitem{innovations}
Montanari, A. and Saberi, A. (2010), \textit{The spread of innovations in social networks}, Proceedings of the National Academy of Sciences, Vol. 107 (47), 20196--20201.


\bibitem{mukherjeeestimation_b}
Mukherjee, S., Son, J. and Bhattacharya, B. (2022), \textit{Estimation in tensor Ising models}, 
{\it Information and Inference: A Journal of the IMA}, Vol. 11 (3), 1457--1500. 

\bibitem{fluctmj}
Mukherjee, S., Son, J. and Bhattacharya, B. (2021), \textit{Fluctuations of the Magnetization in the $p$-Spin Curie–Weiss Model}, 
{\it Communications in Mathematical Physics}, 387, 681--728. 

\bibitem{neykovliu_property}
Neykov, M. and Liu, H. (2019), \textit{Property testing in high-dimensional Ising models}, 
{\it Annals of Statistics}, Vol. 47 (5), 2472--2503.  

\bibitem{oliveira} 
Oliveira, V.M. and Fontanari, J.F. (1997), \textit{Landscape statistics of the $p$-spin Ising model}, Journal of Physics A: Mathematical and General, 30 (24).


\bibitem{wainwright}
Ravikumar, P., Wainwright, M. J. and Lafferty, J. D. (2010), \textit{High-dimensional Ising model selection using $\ell_1$-regularized logistic regression}, Annals of Statistics, 38 (3), 1287--1319.


\bibitem{graphical_models_binary}
Santhanam, N.P., and Wainwright, M.J. (2012), \textit{Information-theoretic limits of selecting binary
graphical models in high dimensions}, IEEE Transactions on Information Theory, Vol. 58 (7), 4117--4134. 




\bibitem{vuffrayint}
Vuffray, M., Misra, S., Lokhov, A.Y., and Chertkov, M. (2016), \textit{Interaction Screening: Efficient and Sample-Optimal
Learning of Ising Models}, arXiv preprint 1605.07252.

\bibitem{wsharp}
Wainwright, M. J. (2009), \textit{Sharp thresholds for high-dimensional and noisy sparsity recovery
using $\ell_1$-constrained quadratic programming (Lasso)}, IEEE Transactions on Information Theory, 55, 2183--2202.

\end{thebibliography}

\end{document}